\documentstyle[twocolumn,epsf,aps]{revtex}

\newcommand{\be}{\begin{equation}}
\newcommand{\ee}{\end{equation}}
\newcommand{\bea}{\begin{eqnarray}}
\newcommand{\eea}{\end{eqnarray}}

\begin{document}

\bibliographystyle{prsty}

\title{
A Non-Equilibrium Defect-Unbinding Transition: 
Defect Trajectories and Loop Statistics 
}

\author{Glen D. Granzow\footnote{current address: Department of Mathematics, Lander University} and Hermann Riecke}

\address{
Department of Engineering Sciences and Applied Mathematics,\\
Northwestern University, 2145 Sheridan Rd, Evanston, IL, 60208, USA
}

\maketitle

\begin{abstract}
 
In a Ginzburg-Landau model for parametrically driven waves a 
transition between a state of ordered and one of disordered 
spatio-temporal defect chaos is found. To characterize the two 
different chaotic states and to get insight into the break-down of the order,
the trajectories of the defects are tracked in detail. Since the defects are
always created and annihilated in pairs the trajectories form loops in space time.
The probability distribution functions for the size of the loops and the number of
defects involved in them undergo a transition from exponential decay in the
ordered regime to a power-law decay in the disordered regime. These 
power laws are also found in a simple lattice model
of randomly created defect pairs that diffuse and annihilate upon collision.

\end{abstract}

\date{\today}



Very early in the investigation of pattern-forming systems far from thermodynamic
equilibrium it 
has been recognized that dislocations in stripe patterns near onset are 
mathematically closely related to defects in the equilibrium 
$xy$-model since both systems are
described by a single complex order parameter $A$ in terms of which the 
dislocations are given by locations of vanishing magnitude $|A|=0$. 
One of the fascinating phase-transition
phenomena studied extensively in the $xy$-model is 
the Kosterlitz-Thouless transition, which is
associated with the unbinding of defect pairs \cite{KoTh73}.  
This motivated early efforts to identify related phenomena 
 also in pattern-forming systems like Rayleigh-B\'enard convection \cite{OcGu83}. 
A direct analogy between these systems
does not hold, however, for a number of reasons. Most significantly, 
the phase transitions occur at finite temperature and
are intimately related to the relevance of fluctuations, whereas in macroscopic systems
like Rayleigh-B\'enard convection the effect of (thermal) noise is negligible 
in most situations \cite{ReHo91}.
Moreover, in the absence of noise there are no
persistent dynamics in the Ginzburg-Landau with real coefficients that describes the 
equilibrium system.

More recently, spatio-temporal chaos in pattern-forming
systems has found considerable attention. A number of variations of convection
experiments have revealed various different types of spatio-temporal 
chaos like spiral-defect chaos (e.g. \cite{MoBo96}), 
domain chaos triggered by the 
K\"uppers-Lortz instability (e.g. \cite{HuPe98}), 
dislocation-dominated chaos in binary-mixture convection \cite{LaSu00} and
in electroconvection \cite{DeAh96}. Extensive 
theoretical investigations have focussed on the complex Ginzburg-Landau 
equation (CGLE) for the complex amplitude of oscillations arising in a Hopf 
bifurcation (e.g. \cite{ChMa96}). 
In contrast to the Ginzburg-Landau equation describing 
the $xy$-model the coefficients in the CGLE are complex and make this
system non-variational. 

Various results have been obtained for the dynamics of defects in 
spatio-temporally chaotic systems. In binary-mixture convection progress has
been made to reconstruct the full wave pattern from the locations of the 
dislocations \cite{LaSu00}. 
The probability distribution function for the number 
of dislocations has been measured in electroconvection \cite{ReRa89} 
and found to agree quite well with results based on the complex 
Ginzburg-Landau 
equation and on a simple diffusive model  \cite{GiLe90}. 
The dynamical relevance of dislocations has been  
demonstrated best so far in work that succeeded in extracting the contribution
of each dislocation to the total fractal 
dimension of the (extensively) chaotic attractor of the CGLE \cite{Eg98}.  

The spatio-temporally chaotic states found in pattern-forming systems typically
arise from ordered states through some transition when a control parameter
is changed. An interesting question is what actually happens when the order
of the pattern breaks down. In analogy
with the melting of two-dimensional crystals \cite{NeHa79,Mi87} one 
may expect that defects in the pattern may play an important role. 
In this Letter we present results for a 
transition between two spatio-temporally chaotic states in a 
Ginzburg-Landau model for parametrically excited waves \cite{GrRi98}. 
While one state is 
disordered in space, the other retains a stripe-like order despite the chaotic
creation and annihilation of defect pairs.  We characterize
the break-down of order in terms of the defect dynamics and find that the 
transition to the disordered state is associated with what one may call an 
unbinding of the pairwise created defects. 
Tentative results for this 
unbinding transition have been presented earlier in \cite{Gr97,GrRi98}.

\begin{figure}
\centerline{\epsfxsize=7cm \epsfbox{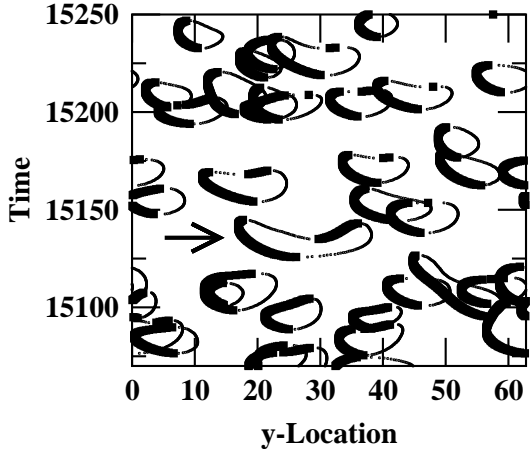} }
\caption{Space-time diagram of defect trajectories.
 for $a=0.25$, $c=-1+4i$, $d=1+0.5i$, $s=0.2$, $g=-1-12i$, and $b=2$. 
Circles and dots denote defects of opposite
topological charge.}
\label{f:loops.stc} 
\end{figure}

To illustrate the possible role of the defect dynamics in the break-down of order,
Fig.\ref{f:loops.stc} presents a space-time diagram of the defect dynamics in the
regime with persistent spatial order.
The $y$-location of each defect with positive topological charge is shown as a 
solid circle
while the $y$-location of each defect with negative charge is shown as a dot.  
The trajectories of almost all of the defect pairs form simple loops in 
space-time. Thus, while in any given system
defects are always annihilated in pairs of opposite charge, in the ordered 
regime  the  annihilating defects have also been created together. 
It is in this sense that we consider them 
to be {\it bound pairs}. The preservation of the stripe-like order 
in this regime can then be understood intuitively, since the dynamics of
defects in such simple defect loops  affect only a very small portion of the 
system and, moreover, render the system almost 
unchanged after their disappearance. In some cases  
the space-time loops involve two (see arrow in fig.\ref{f:loops.stc}) 
or possibly three defect pairs. Then the area of the system that is 
perturbed by the defects is larger,
but after their annihilation the system is still left in essentially the same 
state as before.

The orientation and position of the stripe pattern is affected
significantly only between the defects. Therefore, one may expect that the
destruction of order requires that the defects in a pair 
have to {\em unbind} and separate arbitrarily far from each other. However, it may also 
be sufficient if the defects in a given pair are annihilated by defects from two other pairs, 
which in turn are annihilated by further defect pairs, generating a chain of events whose
trajectory in space-time is a large loop that spans the whole system. Simulations show that
in fact even in the disordered regime most defects are annihilated by their `own' partner 
forming a simple loop and most loops are small compared to the system size. Thus, if a 
distinction between the two regimes can be made based on defect trajectories it must involve
more subtle aspects like the distribution of defect loops as a function of their size.
Using a detailed 
statistical analysis of the defect dynamics  we show in the following 
that indeed the statistical properties
of the defect loops in space-time are qualitatively different in the two regimes.

We consider a Ginzburg-Landau model for parametrically excited waves in an axially
anisotropic system in which at the bifurcation point the waves
travel only along the $x$-axis,
\bea
\partial_t A + s \partial_x A &=& d  \nabla^2 A + a A + b B+\nonumber \\
& &c |A|^2 A + g |B|^2 A, \label{e:A}\\
 \partial_t B - s \partial_x B &=& d^*  \nabla_x^2 B + a^* B
 + b A + \nonumber  \\
& & c^* |B|^2 B + g^* |A|^2B.\label{e:B}
\eea
Here $A$ and $B$ are the complex amplitudes of right- and left-traveling 
waves. All the coefficients except for the group velocity $s$ and the parametric
forcing $b$ are complex. The traveling waves arise in a  Hopf bifurcation at 
$a_r\equiv Re(a)=0$. For $c_r <0$ the bifurcation is supercritical and the 
waves exist for $a_r>0$. 
The parametric forcing  is applied at close to twice the Hopf frequency. It therefore
induces a resonant interaction between the counter-propagating waves at 
linear order \cite{RiCr88,Wa88}.  
For $a_r<0$ and $b^2>a_r^2+a_i^2$ standing waves are excited parametrically that
are phase-locked to the forcing. These waves correspond qualitatively to the
waves excited in Faraday experiments. In all of the simulations we use periodic boundary 
conditions in both directions and solve (\ref{e:A},\ref{e:B}) pseudo-spectrally with a fourth-order
Runge-Kutta scheme using an integrating factor for the linear derivative terms.

Numerical simulations of (\ref{e:A},\ref{e:B}) show two distinct regimes of spatio-temporal
chaos for the parametrically excited waves: 
a conventional regime in which
the spatial correlation function decays rapidly in an essentially isotropic way and a regime
of spatio-temporal chaos which exhibits a strikingly
ordered state in which the correlation function reveals a strong stripe-like 
order that is also apparent 
in individual snapshots \cite{Gr97,GrRi98}. The transition between the two states is discontinuous as
indicated by a jump in the average number of defects. 

In order to get insight into the role of the defect dynamics in this order-disorder 
transition we track each defect from its
creation to its annihilation. In particular, we identify which defect 
was annihilated by which defect. From these data we extract the distribution 
functions for the number of defect pairs that are participating in a space-time
loop and for the
spatial and temporal extent of the loops. Since the relevant information turns out to 
be in the rare, large loops care has to be taken to identify annihilation and creation 
processes reliably and distinguish them from situations in which 
a defect simply moved relatively
far in one time step. In our defect-tracking scheme  we recursively
check the distances between 
defects of equal and of opposite charge. If in consecutive time steps two defects of equal
charge are closer than some threshold value $\delta_1^{(n)}=n \cdot \delta_1$ 
they qualify as a single defect that has
moved from one to the other position. If more than two defects fall into this category
the closest defects are taken to be `continuing' defects. 
Defects that are not continuing defects are candidates for annihilation and creation events.
Among those, two defects of opposite charge and closer than a second threshold 
$\delta_2^{(n)}=n\cdot\delta_2$
are identified as a pair that was annihilated (or created) in this time step. After one
step of this analysis the same analysis is repeated with new, increased
values for the thresholds, $\delta_i^{(n+1)}=(n+1)\cdot \delta_i$, until all defects have
been assigned.  

\begin{figure}
\centerline{\epsfxsize=8cm \epsfbox{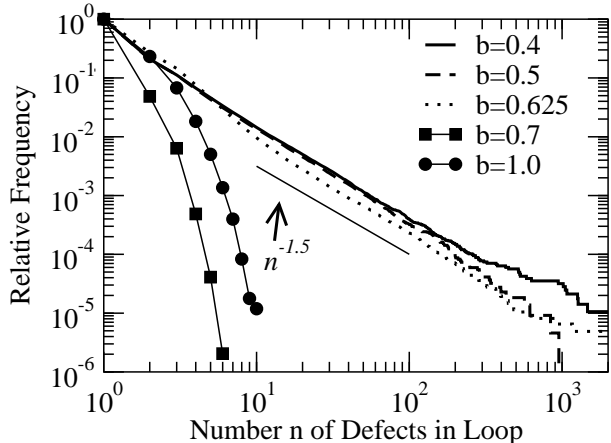} }
\caption{Relative frequency of loops made up of at least $n$ defect pairs. 
Parameters as in
fig.\ref{f:loops.stc} except for $b$. System size $L=1088$ in the disordered regime;
 $L=272$ and $L=136$ for $b=0.7$ and $b=1.0$, respectively.}
\label{f:loops}
\end{figure}

\begin{figure}
\centerline{\epsfxsize=8cm\epsfbox{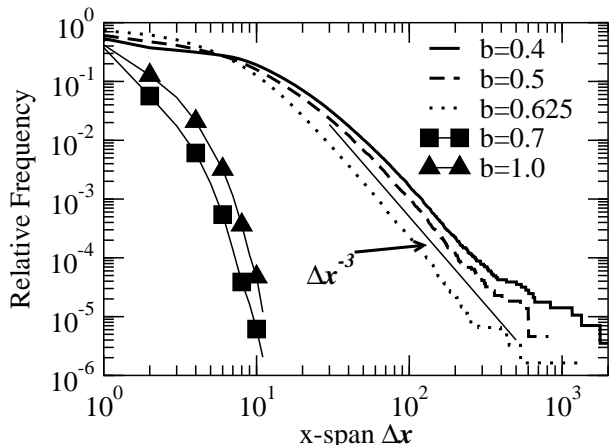} }
\caption{Relative frequency of loops with spatial extent in the $x$-direction 
at least $\Delta x$ . Parameters as in fig.\ref{f:loops}.}
\label{f:xspan}
\end{figure}

\begin{figure}[htb]
\centerline{
\epsfxsize=8cm\epsfbox{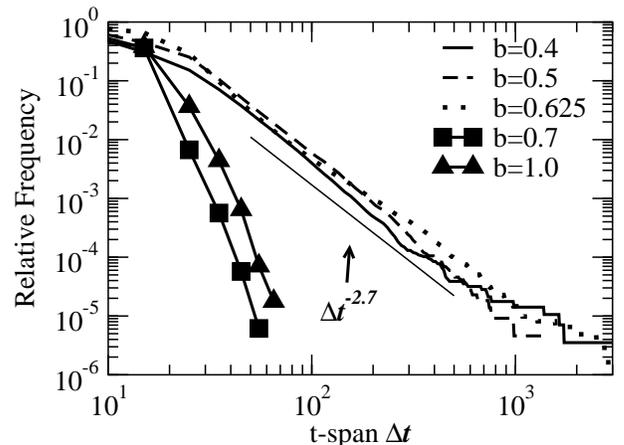}
}
\caption{Relative frequency of loops with lifetime  
at least $\Delta t$. Parameters as in fig.\ref{f:loops}.}
\label{f:tspan}
\end{figure}

Fig.\ref{f:loops} shows the relative frequency of loops consisting of at least 
$n$ defect pairs. The results in figs.\ref{f:loops},\ref{f:xspan},\ref{f:tspan}
are based on 8000 timesteps ($dt=0.5$) with an average number of 7000 defects
at any given time in the disordered regime. In the ordered regime ($b \ge 0.7$)
very few loops contain more
than 5 defect pairs and the distribution decays essentially exponentially. 
In the disordered regime ($b \le 0.625$), however,
 the number  of loops with many 
defects is greatly increased and the decay  of the distribution function
 is only algebraic with an exponent of $\alpha \approx 1.5$.

\typeout{\tt why is there a rise in L=1088 but not in the others? is that because of dt=0.5 instead of dt=0.25?}

Since the defect motion essentially affects only the (vaguely defined)
part of the system between the defects a better indicator for the expected loss of
order are the spatial extents $\Delta x\equiv x_{max}-x_{min}$ and 
$\Delta y\equiv y_{max}-y_{min}$ 
of the defect loops in the $x$- and $y$-direction, respectively. 
Here $x_{min,max}$ and $y_{min,max}$ refer to the minimal and maximal values of $x$ and $y$ 
in any given loop. Note that these values need not be obtained at the same time.
Thus, in principle a small loop could still yield large $\Delta x$ or $\Delta y$ if it traveled.

Similar to fig.\ref{f:loops}, fig.\ref{f:xspan}
shows the relative frequency of loops with size in the $x$-direction
larger than $\Delta x$. 
In the ordered regime the decay is again very rapid and there are 
essentially no loops with $\Delta x$ larger than 10, which is of 
the order of one wavelength.  
This may indicate some pinning of the 
defects by the pattern in that it may restrict their motion to a 
predominantly climbing motion.
 In the disordered pattern,
on the other hand, a noticeable number of loops is of the order 
of the system size  $L=1088$ and even larger 
(i.e. wrapping around the system due to the periodic boundary conditions). 
In a simple interpretation these events would be associated with a persistent 
change in the average wavevector of the pattern. 
 Again, the distribution functions are exponential in 
the ordered regime and exhibit a power law in the disordered regime 
with exponent $\beta \approx 3$. 
The distinction between the regimes is not quite as striking in the
spatial extent $\Delta y$ in the $y$-direction. Even in the ordered regime
the loops can reach a size of $\Delta y \approx 100$, while the loops in the
disordered regime extend to sizes of $\Delta y \approx 1000$. The distribution is 
still exponential in the ordered regime and appears to be power-law in the
disordered regime. However, the measured exponent increases from about
2.8 to 4 as $b$ is increased from 0.4 to 0.625.

The duration $\Delta t$ of the loops is also of interest and the corresponding 
relative frequencies are shown in fig.\ref{f:tspan}. The exponential and power-law
character of the distributions are quite clear in the respective regimes
 with the exponent of the power-law $\gamma \approx 2.7$.

\typeout{\bf test of dependence of parameters in tracking program?}

To give further support for 
the existence of power laws
in the distribution functions in the disordered regime and to get more insight 
into their origin we have investigated
a simple lattice model of the defect dynamics, which is based on the 
observation that the single-defect
statistics in the disordered regime show the same signature as those
of the defect chaos in the CGLE.
 This type of distribution has been shown to arise 
if the defects behave as random walkers that are annihilated upon colliding with 
any other defect of opposite charge \cite{GiLe90}.

\begin{figure}
\centerline{
\epsfxsize=8cm\epsfbox{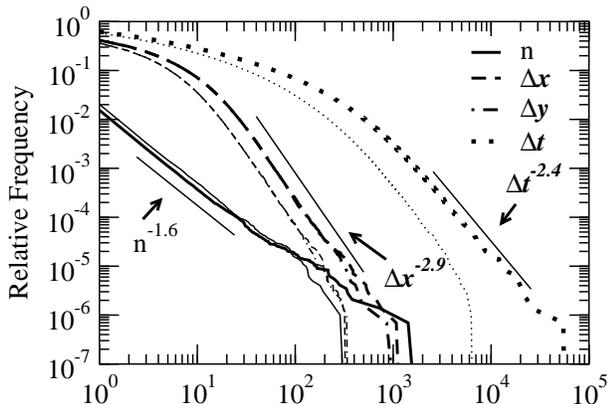} 
}
\caption{Simulations of the lattice model.
Thick lines
denote $p=0.000016$, thin lines $p=0.0001$. $L=1600$.}
\label{f:latticeloops}
\end{figure}

The results of our implementation of the simple lattice model are shown in 
fig.\ref{f:latticeloops}. With a probability $p$, random walkers of opposite 
charge are created pairwise 
at the same randomly chosen site of a 
square lattice. They 
interact only with walkers of opposite charge and annihilate upon contact. 
Fig.\ref{f:latticeloops} shows the loop statistics for these walkers in a 
system of size $L=1600$ and a probability of creation $p=0.000016$ (thick lines)
and $p=0.0001$ (thin lines). All measured 
quantities, i.e. number of defects in a loop and the spatial as well as the
temporal extent of 
the loops, show power-law behavior for large loops. The simplicity of the
lattice model suggests that the power laws are expected to arise in a much wider 
class of spatio-temporally chaotic systems including the CGLE. 
 The exponents are measured to be $\alpha=1.6$, 
$\beta=2.9$, and $\gamma=2.4$, respectively. Thus, even the values of  
exponents agree quite well with those  
 obtained in the simulations of (\ref{e:A},\ref{e:B}).


\typeout{\tt is it really pairs or single defects?} 

In conclusion, using numerical simulations of two coupled complex Ginzburg-Landau
equations we have investigated the break-down of 
 order in a transition from a non-equilibrium, chaotic stripe phase to a
disordered phase. To get insight into the way the order breaks down
we have analyzed the trajectories of the dislocations in the pattern. 
In particular, we have determined the 
statistics of the loops formed in space-time by chains of creation and annihilation 
events 
of oppositely charged defect pairs. The order-disorder transition is related to 
a significant increase in the number of loops that extend over the whole system, which
we associate with the unbinding of defect pairs.
More precisely, the decay of the loop distribution function changes from 
exponential to algebraic in that transition. The algebraic decay is also found
in a simple lattice model of diffusing and annihilating defects, with the
exponents agreeing quite well with those found in the Ginzburg-Landau equations.
The agreement between the Ginzburg-Landau equations and the lattice model
suggests that the power laws may be 
found in a wider class of defect-chaotic
systems and in particular also in the single complex Ginzburg-Landau equation.

The single-defect statistics obtained in the disordered state and in
the lattice model have also been found 
in experiments in electroconvection \cite{ReRa89} and in thermal convection 
in an inclined layer \cite{DaBopriv}. It would be exciting to study also
 the loop distribution functions in these systems. The amount of detailed
data necessary to determine the loop statistics represents presumably 
a challenging task in the current set-ups. 

The simplicity of the lattice model suggests that the power laws may also be 
amenable to an analytic approach. 
A further interesting question is whether the lattice model can be extended
to obtain also a transition from power-law to exponential
decay. Introducing an attractive
interaction between the defects would seem to be a natural choice.
It is not clear whether the interaction should be short- or long-range.
 
HR gratefully acknowledges discussions with H. Chat\'e, M. Cross, 
L. Sander, J. Vi\~nals,
and thanks the Aspen Center for Physics, where parts of this work were performed.
 This work was supported by the Engineering Research Program of the
Office of Basic Energy Sciences at the Department of Energy 
(DE-FG02-92ER14303) and by a grant from NSF (DMS-9804673).


\end{document}